\documentclass[twocolumn,showpacs,preprintnumbers,amsmath,amssymb]{revtex4-1}

\usepackage{graphicx}
\usepackage{dcolumn}
\usepackage{bm}

\begin{document}
\def\3he{$^3$He}
\def\4he{$^4$He}

\title{Flux Bottlenecks in the Mass Superflux in Solid Helium}

\author{Ye. Vekhov}
\author{R.B. Hallock}%
\affiliation{%
Laboratory for Low Temperature Physics, Department of Physics, University of Massachusetts, Amherst, Massachusetts 01003, USA 
}%

\date{\today}

\begin{abstract}
 Superfluid \4he atoms are injected (withdrawn) asymmetrically into (from) a chamber filled with solid \4he. Two \textit{in situ} capacitance pressure gauges are located at the ends of the solid helium sample at different distances from where the superfluid meets the solid \4he. The pressure change kinetics are slower at the more distant gauge. This demonstrates the presence of a mass flux bottleneck \emph{inside} the solid helium sample.  The temperature dependence of the maximum flux reveals a crossover and this is discussed in the context of phase slips on quasi-1-D pathways.
\end{abstract}


\maketitle


The Physics community was greatly stimulated by the possibility of supersolidity suggested by the torsional oscillator experiments of Kim and Chan\cite{Kim2004a,Kim2004b}; many groups reported corroborating evidence.  But, with the realization that previously unexpected shear modulus behavior was present in the solid\cite{Day2007} the community began to question the supersolid interpretation from several perspectives\cite{Reppy2010}.  More recent work has shown that these mechanical effects clearly were dominant and it is now believed that there is little, if any, evidence for a supersolid available from torsional oscillator experiments\cite{Kim2014}.  In conceptually different work, studies of the flux of \4he that passes through a sample cell filled with solid helium have been carried out\cite{Ray2008a,Ray2009b,Ray2010a,Ray2010c,Vekhov2012,Vekhov2014,Vekhov2014b}. These experiments revealed the dependence of the flux rate on the solid helium temperature, the applied chemical potential difference, $\Delta \mu$, and the \3he impurity concentration. They also revealed a dramatic reduction of the flux at a \3he concentration-dependent temperature, $T_d$, a universal temperature dependence above $T_d$ and no flux above $T_h \approx$ 630 mK, etc.   Some of the UMass data\cite{Vekhov2012,Vekhov2014} were interpreted to be consistent with one-dimensional conductivity\cite{Boninsegni2007} through the solid, a so-called Luttinger-like behavior\cite{DelMaestro2010,DelMaestro2011}.

\begin{figure}[htb]
 \centerline{\includegraphics[width=0.8\linewidth,keepaspectratio]{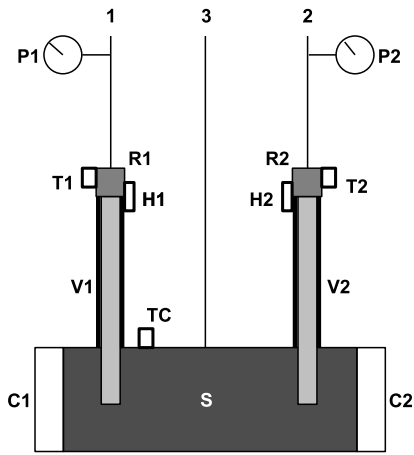}}%
\caption{Schematic diagram of the cell used for
flow experiments.   Two capillaries, 1 and 2, go to
liquid reservoirs $R1$ and $R2$ at the top ends of the Vycor rods, $V1$ and $V2$.  Capillary 3 enters from the side and is used for initial additions of helium to the cell.  Two capacitance
pressure gauges, $C1$ and $C2$, are located on either end of the cell.  Gauges $P1$ and $P2$ are outside the
cryostat.  Heaters $H1$, $H2$, allow the temperatures of the reservoirs to be controlled. For the work reported here the heater $H2$ is used to elevate the temperature of the top of $V2$ to block the flow of atoms to or from reservoir $R2$.  [Reproduced from Fig. 1 in Ref. \cite{Ray2009b}] \label{cell_diagram_bw.eps}}
\end{figure}

In our typical experimental arrangement, referred to as the UMass Sandwich\cite{boris-06} (Fig.~\ref{cell_diagram_bw.eps}), a solid helium sample is located between two superfliud-filled Vycor electrodes, which in turn are connected to two superfluid-filled reservoirs. Recently a number of the key experimental results from the UMass group have been confirmed by Cheng et al.\cite{Balibar.APS2015}. Instead of applying  $\Delta \mu$ between two liquid helium reservoirs at the ends of two Vycor rods, they mechanically squeezed solid helium in one chamber and observed a pressure gauge response in another solid helium chamber which was separated from the other by superfluid-filled Vycor. Based on the data from both groups, Cheng et al. concluded that the flux rate reduction at $T_d$ is likely limited not by solid helium itself but by \3he condensation at the interface between solid helium and liquid helium in the Vycor.  The UMass group had suggested that this flux reduction at $T_d$ might be due to the condensation of \3he atoms at intersections of dislocation cores or on the dislocations themselves\cite{Corboz2008}.  Additional confirmation of evidence for flux through solid helium has been reported recently by Haziot et al.\cite{Chan.APS2015}.

Here we report work that seeks to understand where the limitation of the temperature-dependent mass flux resides for temperatures in the range \emph{above} $T_d$, $T_d < T < T_h$. Is this bottleneck inside the solid helium itself or at the interface between the Vycor and solid helium?  We will show evidence that the bottleneck for $T > T_d$ resides in the solid-filled cell itself and not at the interface between the superfluid in the Vycor and the solid helium.

Solid helium samples are typically grown from the nominally pure (0.17 ppm \3he impurity\cite{Vekhov2014b}) superfluid by filling the sample cell to near the pressure of the melting curve through a direct-access capillary and subsequently increasing the pressure above the melting curve by use of capillaries connected to the sample cell in series with Vycor rods (porous glass with interconnected pores of diameter about 7~nm) at a constant solid helium temperature, $TC \sim 350$~mK. The Vycor rods are 1.40~mm in diameter and 7.62~cm in length. 

In our previous work we created chemical potential differences between two superfluid filled reservoirs to study the resulting flux of helium through the solid-filled cell, Fig.~\ref{cell_diagram_bw.eps}. To explore the bottleneck for $T > T_d$  a modification is made to the experimental arrangement and procedure. Using our standard sample cell\cite{Ray2008a,Ray2009b,Ray2010a,Ray2010b,Vekhov2012}, in this work the top of one Vycor rod, $V2$, is plugged by a high enough temperature  to avoid superfluid mass flux through this rod. So, only Vycor rod $V1$ can be a conduit for \4he atoms. The temperature, $T1$, of the liquid helium reservoir on the top of $V1$ is kept in the rage of $1.46 - 1.51$~K while the solid helium sample (in the form of a horizontal cylinder of $1.84$~cm$^3$ volume and $4.5$~cm length) has a temperature $TC = 0.1-0.8$~K and pressure $25.9 - 26.4$~bar as measured by two \textit{in situ} capacitance pressure gauges, $C1$ and $C2$, located at the ends of the cylindrical sample cell, 10 mm and 33 mm, respectively, from the end of $V1$ in the solid.
Due to the temperature difference between the solid helium and the liquid helium reservoir, $R1$, the thermo-mechanical effect causes a pressure difference between them, $\Delta P$.
The higher $T1$, the larger is the pressure difference between liquid helium in the reservoir at one end of the Vycor rod and the solid helium sample at the other.

Thus, if one wants to change $\Delta P$ and thereby inject (extract) helium atoms into (from) the solid sample, the temperature $T1$ has to be decreased (increased) by some amount, $\delta T$. The response of $C1$ and $C2$ reveals the presence of pressure gradients, if any, and their time-dependent relaxation along the solid helium sample.


There are several possible bottlenecks that might restrict the flow of \4he atoms for $T > T_d$: (a) the interface between the helium in the reservoir and the Vycor rod, (b) the Vycor rod, (c) the interface between the superfluid helium in the Vycor rod and the solid helium, (d) the conduction process in solid helium and (e) the dynamics of edge dislocations which are thought to be responsible for the density changes and pressure changes in the solid helium \cite{Soyler2009}.

As was shown in Ref.~\cite{Vekhov2014}, to avoid a bottleneck anywhere in the Vycor rod and at the interface to the superfluid reservoir the temperature $T1$ typically has to be $< 1.49$~K for a low solid helium temperature, $TC$, where the highest flux rates are observed. It was also shown in  Ref.~\cite{Vekhov2014} that $T1$ can slightly exceed 1.49~K  if $TC > 0.25$~K. We can control the bottleneck due to the Vycor rod by control of $T1$. Next, growth/dissolution of edge dislocations is likely not a bottleneck in the mass flux experiments made earlier \cite{Ray2008a,Ray2009b,Ray2010a,Ray2010b,Vekhov2012,Vekhov2014,Vekhov2014b} because those measurements detected the flux \textit{through} solid helium samples in the presence of a variety of stable pressure gradients inside the solid samples. These various gradients had no  effect on the flux. Thus, we are left here to distinguish between two possibilities, (c) and (d). If the mass flux bottleneck for $T > T_d$ is at the interface between the Vycor rod and the solid helium and not in the solid helium itself,  then we might expect that no difference in the behavior of $C1$ and $C2$ will be observed when atoms enter or leave the solid through $V1$ in response to changes in $T1$. But, if the bottleneck is in the solid helium sample itself, then we might expect that the time dependent behavior of $C1$ and $C2$ will be different due to the difference in the distances of these gauges from the $V1$ rod.

\begin{figure}[htb]
 \centerline{\includegraphics[width=1.2\linewidth,keepaspectratio]{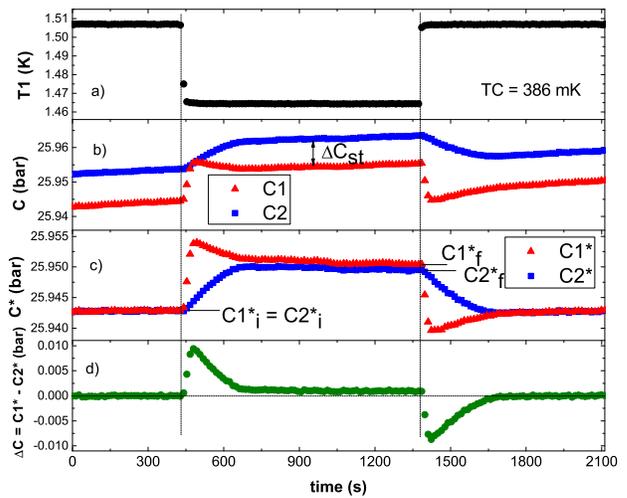}}
\caption{(Color online) Procedure of syringe measurements: a) Liquid helium reservoir temperature, $T1$, change to initiate the mass flux to and from the sample cell at $TC = 386$~mK; b) response of $C1$ and $C2$ \textit{in situ} pressure gauges; c) $C1^*$ and $C2^*$ behavior after pressure drift and steady pressure gradient subtraction; d) the difference between $C1^*$ and $C2^*$.} 
 \label{Fig-Procedure386mK}
\end{figure}


Data that illustrates the procedure of the measurements are shown in
Fig. \ref{Fig-Procedure386mK} for the case of a solid helium temperature, $TC = 386$~mK, midrange in the $T_d < T < T_h$ span in which flux has been previously documented. The time interval between data points is about 13~s. Fig. \ref{Fig-Procedure386mK}~(a) shows a flux-initiating $T1$ temperature change between $1.508 \pm 0.001$ and $1.466 \pm 0.002$~K. One can see that the $T1$ decrease is slower than its increase. We believe that this is due to the thermal impedance between the reservoir $R1$ and the connection to our refrigeration. Increases in $T1$ are prompt because of the proximity of the heater and the reservoir.
The rate of decrease of $T1$ slows at lower solid helium temperatures. During the measurements, the $T1$ temperature change is much faster than the pressure change in solid helium and does not measurably limit the kinetics of this pressure change. Fig. \ref{Fig-Procedure386mK}~(b) presents the response of the \textit{in situ} pressure gauges $C1$ and $C2$. As known from our previous work\cite{Ray2008a,Ray2009b} and can be seen here, $C1$ and $C2$ reveal the presence of a steady pressure difference, $\Delta C_{st}$ (in some samples up to 0.1~bar) in this solid helium sample. This $\Delta C_{st}$ anneals at high enough $TC$ temperatures, usually above 0.7~K. There is a small pressure drift in the solid helium due to a pressure drift in the Vycor filling line which is in turn caused by a level change of liquid helium in the 4K bath of the cryostat.  Fig. \ref{Fig-Procedure386mK}~(c) shows the same $C1$ and $C2$ data, but after subtraction of the pressure drift and steady pressure gradient, now denoted $C1^*$ and $C2^*$, where the * designates that these subtractions have taken place. This background subtraction allows a determination of the difference in the kinetics and steady state behavior after the relaxation of the $C1$ and $C2$. Next, for clarity, we introduce the subscript \emph{i} to denote the value of a parameter before an initial change in the reservoir temperature, $T1$; \emph{f} represents the value of a parameter after a new equilibrium is reached.



In the steady state, there is a small difference between $\Delta C1^* = C1^*_f - C1^*_i$ and $\Delta C2^* = C2^*_f - C2^*_i$ (see Fig. \ref{Fig-Procedure386mK}~(c)) after a decrease in $T1$ and mass addition to the solid.  $\Delta C1^*$ is always larger than $\Delta C2^*$ and this difference is larger for higher $TC$ temperatures.


The kinetic behaviors of $C1$ and $C2$ are significantly different. The pressure gauge closest to the $V1$ Vycor rod, $C1$, shows a much faster response than the further, $C2$, gauge. This leads to the formation and relaxation of a kinetic pressure difference, $\Delta C^*_k(t) = C1^*(t) - C2^*(t)$.  The $C1$ kinetics above  $TC \approx 300$~mK has non-monotonic behavior as shown in Fig.~\ref{Fig-Procedure386mK}(c): a fast increase/decrease and then relaxation back to a new equilibrium value.  The $C2$ kinetics in turn has a monotonic behavior in the whole range of $TC$ temperatures studied. This non-monotonic $C1$ behavior is likely due to a difference between the mass flux conductivity from $V1$ to $C1$ and to $C2$. The flux from $R1$ to $C1$ is faster then from $R1$ to $C2$, apparently due to the difference in distance between $V1$ and the two detectors.

\begin{figure}[htb]
 \centerline{\includegraphics[width=1.1\linewidth,keepaspectratio]{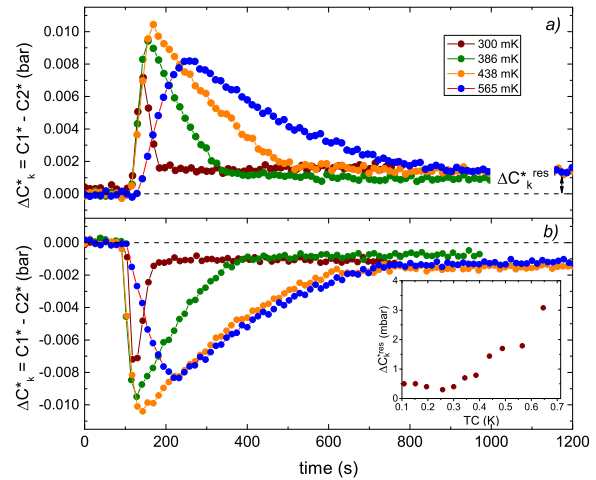}}%
\caption{(Color online) Kinetic pressure difference, $\Delta C^*_k = C1^* - C2^*$, along a solid sample at different $TC$ temperatures after syringing (a) and withdrawing (b) helium atoms to and from the sample cell, respectively. A pressure drift and stationary pressure gradient in a solid sample are subtracted as a background. For the insert, see text discussion.} 
 \label{Fig-dCk}
\end{figure}

Based on data of the sort  presented in Fig.  \ref{Fig-Procedure386mK}(c),  the kinetic pressure
gradient $\Delta C^*_k = C1^* - C2^*$ is shown in Fig.~\ref{Fig-dCk} for several $TC$ temperatures. Thus, Fig.~\ref{Fig-dCk}(a) [(b)] shows $\Delta C_k$ after a $T1$ temperature decrease [increase], i.e. after adding [subtracting] helium atoms through the $V1$ Vycor rod. Several features can be noted here: (a) the higher the $TC$, the slower the $\Delta C_k$ development and its further relaxation; (b) the maximum amplitude of $\Delta C_k$ depends non-monotonically on $TC$; 
(c) $\Delta C_k$ never relaxes to the initial zero-level, furthermore, the higher $TC$, the larger residual $\Delta C_k^{res}$ value (see Fig.~\ref{Fig-dCk}b, insert).

\begin{figure}[htb]
 \centerline{\includegraphics[width=1.0\linewidth,keepaspectratio]{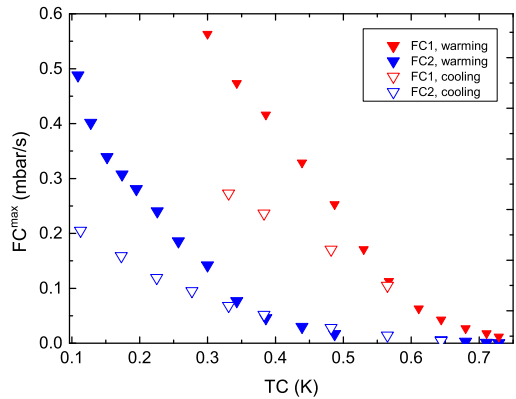}}%
\caption{(Color online) Temperature dependence of the maximum flux (that is not limited by the Vycor) measured by $C1$ (red data points) and $C2$ (blue data points) during step-wise sample warming (solid data points) and subsequent cooling (open data points); $\delta T = 40$~mK.} 
 \label{Fig-FCvsT}
\end{figure}

 In order to determine a measure of the rate of mass flux from $R1$ through the Vycor rod $V1$ to both \textit{in situ} pressure gauges, $C1$ and $C2$, through the solid helium, the derivatives of the pressures $C1$ and $C2$, $FC = dC/dt$, are taken by means of a three-point algorithm and their maximum values for a constant $T1$ change, $\delta T$, are plotted in Fig.~\ref{Fig-FCvsT} for the same solid helium sample. These data are shown for the case of a $T1$ increase.
One can see that $FC1^{max}$ is several times faster than $FC2^{max}$. This $FC$ flux also monotonically decreases with increasing temperature, which is similar to the temperature dependence found for the mass flux \textit{through} a solid helium filled cell and measured by the pressure gauges on the top of both Vycor rods in previous work \cite{Vekhov2012, Vekhov2014b}. The data here are presented for initial sample warming (solid symbols) and subsequent cooling (open symbols). Before warming, the sample had not been annealed. At the highest temperatures the sample starts to anneal resulting in a flux rate decrease seen on cooling.

\begin{figure}[htb]
 \centerline{\includegraphics[width=1.1\linewidth,keepaspectratio]{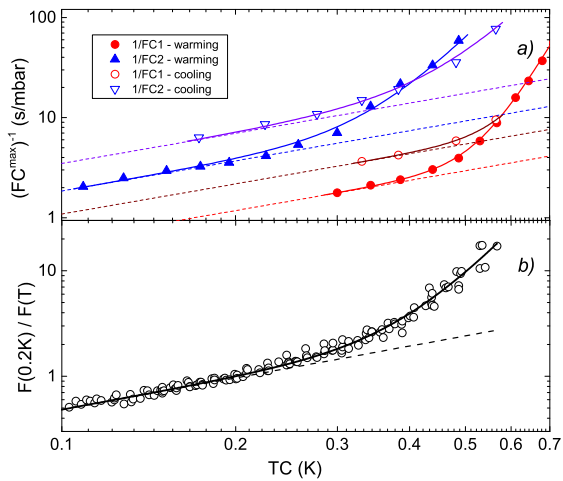}}%
\caption{(Color online) (a) Temperature dependence of the flux resistance, $(FC^{max})^{-1}$, measured by $C1$ (circles) and $C2$ (triangles); $\delta T = 40$~mK. Although the data are limited, the solid lines represent fits to $AT + BT^\alpha$; $\alpha$ is not particularly well determined in the range 5-10.  (b) With flux resistance data from many samples from Ref.\cite{Vekhov2014b}, $\alpha = 5.8 \pm 0.3$.  The dashed lines in all cases represent a linear temperature dependence. } 
 \label{Fig-1FCvslogT}
\end{figure}

 The data from Fig.~\ref{Fig-FCvsT} can be inverted, $(FC^{max})^{-1}$,  to obtain something we might call a flux resistance as shown in Fig.~\ref{Fig-1FCvslogT}(a). When this is done, it appears that there is a crossover in behavior from one conductance regime to another.
We note here that our recent mass flux data  \textit{through} many solid samples of various \3he concentrations for $T>T_d$ (see Ref.\cite{Vekhov2014b}, Fig.4) demonstrated a universal temperature dependence for the maximum flux, $F$, when normalized, $F/F(T=0.2K)$.   That same data can also be  presented as a flux resistance, $F(T=0.2K)/F$, as shown in Fig.~\ref{Fig-1FCvslogT}(b). When that is done we see that it demonstrates a very similar behavior. That is, the flux measured \emph{through} many samples also shows this apparent crossover behavior.  



It is a question for further study to determine what the origin of the contributions to the temperature dependence of the conductance is that gives rise to this apparent crossover behavior.  It is possible that quantum phase slips\cite{Zaikin2008} are responsible, as has been seen for narrow superconducting fibers\cite{Tian2005,Pai2008,Zaikin2008}.

In conclusion we have shown that measured changes in the pressure in a cell filled with solid helium when mass is injected depend on the relative location of the pressure gauges with which the measurements are made.  We interpret these measurements to indicate that the bottleneck to the temperature dependent flux \emph{above} the temperature $T_d$ is due to superflow dissipation in the solid sample itself and the flux is not limited by interface effects where superfluid in the Vycor contacts the solid helium; the bottleneck for $T > T_d$ is a bulk phenomena and not a surface effect. Our results support the possibility that the dissipative mass flux seen may be due to the superfluid cores of edge dislocations that carry the flux, but do not conclusively prove it.




We appreciate the early contributions to the apparatus by M.W. Ray and conversations with colleagues in the field, particularly B.V. Svistunov, J. Beamish and M. Chan. This work was supported by the National Science Foundation grant No. DMR 12-05217  and by Trust Funds administered by the Univ. of Mass. Amherst.

\bibliography{ref3a}

\begin{thebibliography}{24}%
\makeatletter
\providecommand \@ifxundefined [1]{%
 \@ifx{#1\undefined}
}%
\providecommand \@ifnum [1]{%
 \ifnum #1\expandafter \@firstoftwo
 \else \expandafter \@secondoftwo
 \fi
}%
\providecommand \@ifx [1]{%
 \ifx #1\expandafter \@firstoftwo
 \else \expandafter \@secondoftwo
 \fi
}%
\providecommand \natexlab [1]{#1}%
\providecommand \enquote  [1]{``#1''}%
\providecommand \bibnamefont  [1]{#1}%
\providecommand \bibfnamefont [1]{#1}%
\providecommand \citenamefont [1]{#1}%
\providecommand \href@noop [0]{\@secondoftwo}%
\providecommand \href [0]{\begingroup \@sanitize@url \@href}%
\providecommand \@href[1]{\@@startlink{#1}\@@href}%
\providecommand \@@href[1]{\endgroup#1\@@endlink}%
\providecommand \@sanitize@url [0]{\catcode `\\12\catcode `\$12\catcode
  `\&12\catcode `\#12\catcode `\^12\catcode `\_12\catcode `\%12\relax}%
\providecommand \@@startlink[1]{}%
\providecommand \@@endlink[0]{}%
\providecommand \url  [0]{\begingroup\@sanitize@url \@url }%
\providecommand \@url [1]{\endgroup\@href {#1}{\urlprefix }}%
\providecommand \urlprefix  [0]{URL }%
\providecommand \Eprint [0]{\href }%
\providecommand \doibase [0]{http://dx.doi.org/}%
\providecommand \selectlanguage [0]{\@gobble}%
\providecommand \bibinfo  [0]{\@secondoftwo}%
\providecommand \bibfield  [0]{\@secondoftwo}%
\providecommand \translation [1]{[#1]}%
\providecommand \BibitemOpen [0]{}%
\providecommand \bibitemStop [0]{}%
\providecommand \bibitemNoStop [0]{.\EOS\space}%
\providecommand \EOS [0]{\spacefactor3000\relax}%
\providecommand \BibitemShut  [1]{\csname bibitem#1\endcsname}%
\let\auto@bib@innerbib\@empty
\bibitem [{\citenamefont {Kim}\ and\ \citenamefont
  {Chan}(2004{\natexlab{a}})}]{Kim2004a}%
  \BibitemOpen
  \bibfield  {author} {\bibinfo {author} {\bibfnamefont {E.}~\bibnamefont
  {Kim}}\ and\ \bibinfo {author} {\bibfnamefont {M.}~\bibnamefont {Chan}},\
  }\href@noop {} {\bibfield  {journal} {\bibinfo  {journal} {Nature}\ }\textbf
  {\bibinfo {volume} {427}},\ \bibinfo {pages} {225} (\bibinfo {year}
  {2004}{\natexlab{a}})}\BibitemShut {NoStop}%
\bibitem [{\citenamefont {Kim}\ and\ \citenamefont
  {Chan}(2004{\natexlab{b}})}]{Kim2004b}%
  \BibitemOpen
  \bibfield  {author} {\bibinfo {author} {\bibfnamefont {E.}~\bibnamefont
  {Kim}}\ and\ \bibinfo {author} {\bibfnamefont {M.}~\bibnamefont {Chan}},\
  }\href@noop {} {\bibfield  {journal} {\bibinfo  {journal} {Science}\ }\textbf
  {\bibinfo {volume} {305}},\ \bibinfo {pages} {1941} (\bibinfo {year}
  {2004}{\natexlab{b}})}\BibitemShut {NoStop}%
\bibitem [{\citenamefont {Day}\ and\ \citenamefont {Beamish}(2007)}]{Day2007}%
  \BibitemOpen
  \bibfield  {author} {\bibinfo {author} {\bibfnamefont {J.}~\bibnamefont
  {Day}}\ and\ \bibinfo {author} {\bibfnamefont {J.}~\bibnamefont {Beamish}},\
  }\href {\doibase 10.1038/nature06383} {\bibfield  {journal} {\bibinfo
  {journal} {Nature}\ }\textbf {\bibinfo {volume} {450}},\ \bibinfo {pages}
  {853} (\bibinfo {year} {2007})}\BibitemShut {NoStop}%
\bibitem [{\citenamefont {Reppy}(2010)}]{Reppy2010}%
  \BibitemOpen
  \bibfield  {author} {\bibinfo {author} {\bibfnamefont {J.~D.}\ \bibnamefont
  {Reppy}},\ }\href {\doibase 10.1103/PhysRevLett.104.255301} {\bibfield
  {journal} {\bibinfo  {journal} {Phys. Rev. Lett.}\ }\textbf {\bibinfo
  {volume} {104}},\ \bibinfo {pages} {255301} (\bibinfo {year}
  {2010})}\BibitemShut {NoStop}%
\bibitem [{\citenamefont {Kim}\ and\ \citenamefont {Chan}(2014)}]{Kim2014}%
  \BibitemOpen
  \bibfield  {author} {\bibinfo {author} {\bibfnamefont {D.~Y.}\ \bibnamefont
  {Kim}}\ and\ \bibinfo {author} {\bibfnamefont {M.~H.~W.}\ \bibnamefont
  {Chan}},\ }\href@noop {} {\bibfield  {journal} {\bibinfo  {journal} {Phys.
  Rev. B}\ }\textbf {\bibinfo {volume} {90}},\ \bibinfo {pages} {064503}
  (\bibinfo {year} {2014})}\BibitemShut {NoStop}%
\bibitem [{\citenamefont {Ray}\ and\ \citenamefont {Hallock}(2008)}]{Ray2008a}%
  \BibitemOpen
  \bibfield  {author} {\bibinfo {author} {\bibfnamefont {M.~W.}\ \bibnamefont
  {Ray}}\ and\ \bibinfo {author} {\bibfnamefont {R.~B.}\ \bibnamefont
  {Hallock}},\ }\href {\doibase 10.1103/PhysRevLett.100.235301} {\bibfield
  {journal} {\bibinfo  {journal} {Phys. Rev. Lett.}\ }\textbf {\bibinfo
  {volume} {100}},\ \bibinfo {eid} {235301} (\bibinfo {year}
  {2008})}\BibitemShut {NoStop}%
\bibitem [{\citenamefont {Ray}\ and\ \citenamefont {Hallock}(2009)}]{Ray2009b}%
  \BibitemOpen
  \bibfield  {author} {\bibinfo {author} {\bibfnamefont {M.~W.}\ \bibnamefont
  {Ray}}\ and\ \bibinfo {author} {\bibfnamefont {R.~B.}\ \bibnamefont
  {Hallock}},\ }\href {\doibase 10.1103/PhysRevB.79.224302} {\bibfield
  {journal} {\bibinfo  {journal} {Phys. Rev. B}\ }\textbf {\bibinfo {volume}
  {79}},\ \bibinfo {eid} {224302} (\bibinfo {year} {2009})}\BibitemShut
  {NoStop}%
\bibitem [{\citenamefont {Ray}\ and\ \citenamefont
  {Hallock}(2010{\natexlab{a}})}]{Ray2010a}%
  \BibitemOpen
  \bibfield  {author} {\bibinfo {author} {\bibfnamefont {M.~W.}\ \bibnamefont
  {Ray}}\ and\ \bibinfo {author} {\bibfnamefont {R.~B.}\ \bibnamefont
  {Hallock}},\ }\href {\doibase 10.1103/PhysRevB.81.214523} {\bibfield
  {journal} {\bibinfo  {journal} {Phys. Rev. B}\ }\textbf {\bibinfo {volume}
  {81}},\ \bibinfo {pages} {214523} (\bibinfo {year}
  {2010}{\natexlab{a}})}\BibitemShut {NoStop}%
\bibitem [{\citenamefont {Ray}\ and\ \citenamefont
  {Hallock}(2010{\natexlab{b}})}]{Ray2010c}%
  \BibitemOpen
  \bibfield  {author} {\bibinfo {author} {\bibfnamefont {M.~W.}\ \bibnamefont
  {Ray}}\ and\ \bibinfo {author} {\bibfnamefont {R.~B.}\ \bibnamefont
  {Hallock}},\ }\href {\doibase 10.1103/PhysRevLett.105.145301} {\bibfield
  {journal} {\bibinfo  {journal} {Phys. Rev. Lett.}\ }\textbf {\bibinfo
  {volume} {105}},\ \bibinfo {pages} {145301} (\bibinfo {year}
  {2010}{\natexlab{b}})}\BibitemShut {NoStop}%
\bibitem [{\citenamefont {Vekhov}\ and\ \citenamefont
  {Hallock}(2012)}]{Vekhov2012}%
  \BibitemOpen
  \bibfield  {author} {\bibinfo {author} {\bibfnamefont {Y.}~\bibnamefont
  {Vekhov}}\ and\ \bibinfo {author} {\bibfnamefont {R.~B.}\ \bibnamefont
  {Hallock}},\ }\href@noop {} {\bibfield  {journal} {\bibinfo  {journal} {Phys.
  Rev. Lett.}\ }\textbf {\bibinfo {volume} {109}},\ \bibinfo {pages} {045303}
  (\bibinfo {year} {2012})}\BibitemShut {NoStop}%
\bibitem [{\citenamefont {Vekhov}\ and\ \citenamefont
  {Hallock}(2014)}]{Vekhov2014}%
  \BibitemOpen
  \bibfield  {author} {\bibinfo {author} {\bibfnamefont {Y.}~\bibnamefont
  {Vekhov}}\ and\ \bibinfo {author} {\bibfnamefont {R.~B.}\ \bibnamefont
  {Hallock}},\ }\href@noop {} {\bibfield  {journal} {\bibinfo  {journal} {Phys.
  Rev. B}\ }\textbf {\bibinfo {volume} {90}},\ \bibinfo {pages} {134511}
  (\bibinfo {year} {2014})}\BibitemShut {NoStop}%
\bibitem [{\citenamefont {Vekhov}\ \emph {et~al.}(2014)\citenamefont {Vekhov},
  \citenamefont {Mullin},\ and\ \citenamefont {Hallock}}]{Vekhov2014b}%
  \BibitemOpen
  \bibfield  {author} {\bibinfo {author} {\bibfnamefont {Y.}~\bibnamefont
  {Vekhov}}, \bibinfo {author} {\bibfnamefont {W.~J.}\ \bibnamefont {Mullin}},
  \ and\ \bibinfo {author} {\bibfnamefont {R.~B.}\ \bibnamefont {Hallock}},\
  }\href@noop {} {\bibfield  {journal} {\bibinfo  {journal} {Phys. Rev. Lett.}\
  }\textbf {\bibinfo {volume} {113}},\ \bibinfo {pages} {035302} (\bibinfo
  {year} {2014})}\BibitemShut {NoStop}%
\bibitem [{\citenamefont {Boninsegni}\ \emph {et~al.}(2007)\citenamefont
  {Boninsegni}, \citenamefont {Kuklov}, \citenamefont {Pollet}, \citenamefont
  {Prokof'ev}, \citenamefont {Svistunov},\ and\ \citenamefont
  {Troyer}}]{Boninsegni2007}%
  \BibitemOpen
  \bibfield  {author} {\bibinfo {author} {\bibfnamefont {M.}~\bibnamefont
  {Boninsegni}}, \bibinfo {author} {\bibfnamefont {A.~B.}\ \bibnamefont
  {Kuklov}}, \bibinfo {author} {\bibfnamefont {L.}~\bibnamefont {Pollet}},
  \bibinfo {author} {\bibfnamefont {N.~V.}\ \bibnamefont {Prokof'ev}}, \bibinfo
  {author} {\bibfnamefont {B.~V.}\ \bibnamefont {Svistunov}}, \ and\ \bibinfo
  {author} {\bibfnamefont {M.}~\bibnamefont {Troyer}},\ }\href {\doibase
  10.1103/PhysRevLett.99.035301} {\bibfield  {journal} {\bibinfo  {journal}
  {Phys. Rev. Lett.}\ }\textbf {\bibinfo {volume} {99}},\ \bibinfo {eid}
  {035301} (\bibinfo {year} {2007})}\BibitemShut {NoStop}%
\bibitem [{\citenamefont {Del~Maestro}\ and\ \citenamefont
  {Affleck}(2010)}]{DelMaestro2010}%
  \BibitemOpen
  \bibfield  {author} {\bibinfo {author} {\bibfnamefont {A.}~\bibnamefont
  {Del~Maestro}}\ and\ \bibinfo {author} {\bibfnamefont {I.}~\bibnamefont
  {Affleck}},\ }\href {\doibase 10.1103/PhysRevB.82.060515} {\bibfield
  {journal} {\bibinfo  {journal} {Phys. Rev. B}\ }\textbf {\bibinfo {volume}
  {82}},\ \bibinfo {eid} {060515(R)} (\bibinfo {year} {2010})}\BibitemShut
  {NoStop}%
\bibitem [{\citenamefont {Del~Maestro}\ \emph {et~al.}(2011)\citenamefont
  {Del~Maestro}, \citenamefont {Boninsegni},\ and\ \citenamefont
  {Affleck}}]{DelMaestro2011}%
  \BibitemOpen
  \bibfield  {author} {\bibinfo {author} {\bibfnamefont {A.}~\bibnamefont
  {Del~Maestro}}, \bibinfo {author} {\bibfnamefont {M.}~\bibnamefont
  {Boninsegni}}, \ and\ \bibinfo {author} {\bibfnamefont {I.}~\bibnamefont
  {Affleck}},\ }\href {\doibase 10.1103/PhysRevLett.106.105303} {\bibfield
  {journal} {\bibinfo  {journal} {Phys. Rev. Lett.}\ }\textbf {\bibinfo
  {volume} {106}},\ \bibinfo {eid} {105303} (\bibinfo {year}
  {2011})}\BibitemShut {NoStop}%
\bibitem [{\citenamefont {Svistunov}(2006)}]{boris-06}%
  \BibitemOpen
  \bibfield  {author} {\bibinfo {author} {\bibfnamefont {B.}~\bibnamefont
  {Svistunov}},\ }\href
  {http://online.itp.ucsb.edu/online/smatter-m06/svistunov/} {\enquote
  {\bibinfo {title}
  {http://online.itp.ucsb.edu/online/smatter-m06/svistunov/},}\ } (\bibinfo
  {year} {2006})\BibitemShut {NoStop}%
\bibitem [{\citenamefont {Cheng}\ \emph {et~al.}()\citenamefont {Cheng},
  \citenamefont {Beamish}, \citenamefont {Fefferman}, \citenamefont {Souris},\
  and\ \citenamefont {Balibar}}]{Balibar.APS2015}%
  \BibitemOpen
  \bibfield  {author} {\bibinfo {author} {\bibfnamefont {Z.}~\bibnamefont
  {Cheng}}, \bibinfo {author} {\bibfnamefont {J.}~\bibnamefont {Beamish}},
  \bibinfo {author} {\bibfnamefont {A.}~\bibnamefont {Fefferman}}, \bibinfo
  {author} {\bibfnamefont {F.}~\bibnamefont {Souris}}, \ and\ \bibinfo {author}
  {\bibfnamefont {S.}~\bibnamefont {Balibar}},\ }in\ \href@noop {} {\emph
  {\bibinfo {booktitle} {APS March Meeting 2015, March 2 - 6, San Antonio,
  Texas}}}\BibitemShut {NoStop}%
\bibitem [{\citenamefont {Corboz}\ \emph {et~al.}(2008)\citenamefont {Corboz},
  \citenamefont {Pollet}, \citenamefont {Prokof'ev},\ and\ \citenamefont
  {Troyer}}]{Corboz2008}%
  \BibitemOpen
  \bibfield  {author} {\bibinfo {author} {\bibfnamefont {P.}~\bibnamefont
  {Corboz}}, \bibinfo {author} {\bibfnamefont {L.}~\bibnamefont {Pollet}},
  \bibinfo {author} {\bibfnamefont {N.~V.}\ \bibnamefont {Prokof'ev}}, \ and\
  \bibinfo {author} {\bibfnamefont {M.}~\bibnamefont {Troyer}},\ }\href@noop {}
  {\bibfield  {journal} {\bibinfo  {journal} {Phys. Rev. Lett.}\ }\textbf
  {\bibinfo {volume} {101}},\ \bibinfo {pages} {155302} (\bibinfo {year}
  {2008})}\BibitemShut {NoStop}%
\bibitem [{\citenamefont {Haziot}\ \emph {et~al.}()\citenamefont {Haziot},
  \citenamefont {Kim},\ and\ \citenamefont {Chan}}]{Chan.APS2015}%
  \BibitemOpen
  \bibfield  {author} {\bibinfo {author} {\bibfnamefont {A.}~\bibnamefont
  {Haziot}}, \bibinfo {author} {\bibfnamefont {D.}~\bibnamefont {Kim}}, \ and\
  \bibinfo {author} {\bibfnamefont {M.}~\bibnamefont {Chan}},\ }in\ \href@noop
  {} {\emph {\bibinfo {booktitle} {APS March Meeting 2015, March 2 - 6, San
  Antonio, Texas}}}\BibitemShut {NoStop}%
\bibitem [{\citenamefont {Ray}\ and\ \citenamefont
  {Hallock}(2010{\natexlab{c}})}]{Ray2010b}%
  \BibitemOpen
  \bibfield  {author} {\bibinfo {author} {\bibfnamefont {M.~W.}\ \bibnamefont
  {Ray}}\ and\ \bibinfo {author} {\bibfnamefont {R.~B.}\ \bibnamefont
  {Hallock}},\ }\href {\doibase 10.1103/PhysRevB.82.012502} {\bibfield
  {journal} {\bibinfo  {journal} {Phys. Rev. B}\ }\textbf {\bibinfo {volume}
  {82}},\ \bibinfo {pages} {012502} (\bibinfo {year}
  {2010}{\natexlab{c}})}\BibitemShut {NoStop}%
\bibitem [{\citenamefont {\c{S} G.~S\"{o}yler}\ \emph
  {et~al.}(2009)\citenamefont {\c{S} G.~S\"{o}yler}, \citenamefont {Kuklov},
  \citenamefont {Pollet}, \citenamefont {Prokof'ev},\ and\ \citenamefont
  {Svistunov}}]{Soyler2009}%
  \BibitemOpen
  \bibfield  {author} {\bibinfo {author} {\bibnamefont {\c{S} G.~S\"{o}yler}},
  \bibinfo {author} {\bibfnamefont {A.~B.}\ \bibnamefont {Kuklov}}, \bibinfo
  {author} {\bibfnamefont {L.}~\bibnamefont {Pollet}}, \bibinfo {author}
  {\bibfnamefont {N.~V.}\ \bibnamefont {Prokof'ev}}, \ and\ \bibinfo {author}
  {\bibfnamefont {B.~V.}\ \bibnamefont {Svistunov}},\ }\href {\doibase
  10.1103/PhysRevLett.103.175301} {\bibfield  {journal} {\bibinfo  {journal}
  {Phys. Rev. Lett.}\ }\textbf {\bibinfo {volume} {103}},\ \bibinfo {eid}
  {175301} (\bibinfo {year} {2009})}\BibitemShut {NoStop}%
\bibitem [{\citenamefont {Arutyunov}\ \emph {et~al.}(2008)\citenamefont
  {Arutyunov}, \citenamefont {Golubev},\ and\ \citenamefont
  {Zaikin}}]{Zaikin2008}%
  \BibitemOpen
  \bibfield  {author} {\bibinfo {author} {\bibfnamefont {K.}~\bibnamefont
  {Arutyunov}}, \bibinfo {author} {\bibfnamefont {D.}~\bibnamefont {Golubev}},
  \ and\ \bibinfo {author} {\bibfnamefont {A.}~\bibnamefont {Zaikin}},\
  }\href@noop {} {\bibfield  {journal} {\bibinfo  {journal} {Physics Reports}\
  }\textbf {\bibinfo {volume} {464}},\ \bibinfo {pages} {1} (\bibinfo {year}
  {2008})}\BibitemShut {NoStop}%
\bibitem [{\citenamefont {Tian}\ \emph {et~al.}(2005)\citenamefont {Tian},
  \citenamefont {Wang}, \citenamefont {Kurtz}, \citenamefont {Liu},
  \citenamefont {Chan}, \citenamefont {Mayer},\ and\ \citenamefont
  {Mallouk}}]{Tian2005}%
  \BibitemOpen
  \bibfield  {author} {\bibinfo {author} {\bibfnamefont {M.}~\bibnamefont
  {Tian}}, \bibinfo {author} {\bibfnamefont {J.}~\bibnamefont {Wang}}, \bibinfo
  {author} {\bibfnamefont {J.~S.}\ \bibnamefont {Kurtz}}, \bibinfo {author}
  {\bibfnamefont {Y.}~\bibnamefont {Liu}}, \bibinfo {author} {\bibfnamefont
  {M.~H.~W.}\ \bibnamefont {Chan}}, \bibinfo {author} {\bibfnamefont {T.~S.}\
  \bibnamefont {Mayer}}, \ and\ \bibinfo {author} {\bibfnamefont {T.~E.}\
  \bibnamefont {Mallouk}},\ }\href@noop {} {\bibfield  {journal} {\bibinfo
  {journal} {Phys. Rev. B}\ }\textbf {\bibinfo {volume} {71}},\ \bibinfo
  {pages} {104521} (\bibinfo {year} {2005})}\BibitemShut {NoStop}%
\bibitem [{\citenamefont {Pai}\ \emph {et~al.}(2008)\citenamefont {Pai},
  \citenamefont {Shimshoni},\ and\ \citenamefont {Andrei}}]{Pai2008}%
  \BibitemOpen
  \bibfield  {author} {\bibinfo {author} {\bibfnamefont {G.~V.}\ \bibnamefont
  {Pai}}, \bibinfo {author} {\bibfnamefont {E.}~\bibnamefont {Shimshoni}}, \
  and\ \bibinfo {author} {\bibfnamefont {N.}~\bibnamefont {Andrei}},\
  }\href@noop {} {\bibfield  {journal} {\bibinfo  {journal} {Phys. Rev. B}\
  }\textbf {\bibinfo {volume} {77}},\ \bibinfo {pages} {104528} (\bibinfo
  {year} {2008})}\BibitemShut {NoStop}%
\end{thebibliography}%

\end{document}